# Anti-Stokes Photoluminescence from CsPbBr$_3$ Nanostructures Embedded in a Cs$_4$PbBr$_6$ Crystal

*Yuto Kajino,[1] Shuji Otake,[1] Takumi Yamada,[2] Kazunobu Kojima,[3] Tomoya Nakamura,[2] Atsushi, Wakamiya,[2] Yoshihiko Kanemitsu,[2] and Yasuhiro Yamada[1,*]*

[1] Graduate School of Science, Chiba University, Inage, Chiba 263-8522, Japan

[2] Institute for Chemical Research, Kyoto University, Gokasho, Uji 611-0011, Japan

[3] Institute of Multidisciplinary Research for Advanced Materials, Tohoku University, Aoba, Sendai 980-8577, Japan

**ABSTRACT**

Lead halide perovskites possess high photoluminescence (PL) efficiency and strong electron-phonon interactions, and therefore the optical cooling using up-conversion PL has been expected. We investigate anti-Stokes PL from green-luminescent Cs$_4$PbBr$_6$, whose origin is attributable to CsPbBr$_3$ nanostructures embedded in a Cs$_4$PbBr$_6$ crystal. Because of the high transparency, low refractive index, and high stability of Cs$_4$PbBr$_6$, the green PL displays high external quantum efficiency without photo-degradation. Time-resolved PL spectroscopy reveals the excitonic behaviors in recombination process. The shape of the PL spectrum is almost independent of excitation photon energy, which means that the spectral width is determined by homogeneous broadening. We demonstrate that the phonon-assisted process dominates the Urbach tail of optical absorption and anti-Stokes PL at room temperature. Anti-Stokes PL is observed down to 70 K. We determine the temperature dependence of the Urbach energy and estimate the strength of the electron-phonon coupling. Our spectroscopic data show that CsPbBr$_3$ nanostructures have potentially useful features for optical cooling.

Corresponding author: * yasuyamada@chiba-u.jp

Optical refrigeration has been proposed as a technique for cooling materials using anti-Stokes photoluminescence (PL) [1]. Anti-Stokes light emission is the up-converted PL that emits light with higher energy than the excitation light. If the anti-Stokes PL efficiency becomes sufficiently high, light emission reduces the energy of the system, resulting in cooling. Solid-state optical refrigeration can offer vibration-free, non-contact, and small cooling devices [2,3]. While glasses or crystals doped with rare-earth ions have been intensively studied [2,4], semiconductor optical refrigeration has been expected because of the strong optical absorption and possible low-temperature cooling [5–7]. In 2013, Zhang *et al.* reported optical cooling of CdS nanobelts [8]. This result has stimulated the research of optical refrigeration and led to debates regarding cooling mechanisms [9–11]. Since the phonon-assisted anti-Stokes PL is the essential process for semiconductor optical cooling, it is considered that the materials combining strong electron-phonon coupling and high PL quantum efficiency are suitable for optical refrigeration.

In this context, halide perovskites emerged as new candidate materials for semiconductor optical refrigeration [12–16]. Halide perovskites have been subject to extensive research for solar cells, light-emitting diodes, and various optoelectronic applications [17–22]. These perovskites exhibit excellent optical and electronic properties, such as strong band-to-band optical transitions and high PL efficiencies [12,23–27]. Efficient anti-Stokes PL and strong electron–phonon interactions have also been reported, suggesting their potential for optical refrigeration [13–15, 28,29]. In fact, some research groups have reported net cooling in micrometer- or nano-sized perovskites such as $CH_3NH_3PbI_3$, $(C_6H_5C_2H_4NH_3)_2PbI_4$, and $CsPbBr_3$ [13,16]. However, the validity of optical cooling experiments is still controversial [10,11]. Moreover, conventional halide perovskites have a problem in their stability; their PL efficiency decreases under exposure to air and continuous light irradiation [30–32]. This problem complicates the realization of optical cooling with halide perovskites as well as the interpretation of cooling experiments.

In this study, we focus on $Cs_4PbBr_6$ bulk crystals that exhibit efficient green PL. While the origin of the strong green PL has been discussed so far [33-36], recent studies revealed that the origin of the green PL is CsPbBr$_3$ nanostructures embedded in a Cs$_4$PbBr$_6$ crystal (see Figure 1(a)) [36–38], which is also supported by our spectroscopic data (see Supplemental Material for more detail) [39]. Accordingly, we hereafter refer to the green-luminescent Cs$_4$PbBr$_6$ crystal as a CsPbBr$_3$/Cs$_4$PbBr$_6$ crystal. This perovskite composite material has extremely high PL quantum efficiencies (> 97% [34]) and high photostability [36]. The high transparency and low refractive index of Cs$_4$PbBr$_6$ is advantageous for efficient light extraction [50]. Here, we discuss the physical mechanisms of the up-conversion PL of such guest/host perovskite composite crystals: we confirm a strong anti-Stokes PL and find that the shape of the PL spectrum is almost independent of excitation photon energy, which means that the spectral width is determined by homogeneous broadening. Temperature-dependence of the optical spectra reveals that the Urbach tail of absorption and the anti-Stokes PL are dominated by intrinsic phonon-assisted process. We consider these remarkable features from the viewpoint of electron–phonon interactions and describe the potential of halide perovskite nanostructures for optical cooling.

We fabricated CsPbBr$_3$/Cs$_4$PbBr$_6$ crystals by solution process. The detailed growth techniques were described elsewhere [35]. In short, CsBr and PbBr$_2$ powders with a 3:1 molar ratio were dissolved with DMF and HBr mixed solution (5:4 volume ratio) at 90 °C. The solution was gradually cooled down to about 50 °C in several days, and we obtained green-emissive crystals with a few millimeters in size. We used as-grown crystals in our measurements and confirmed their crystal structure by X-ray diffraction measurement [39]. All measurements except those for Figure 5 were performed at room temperature.

Intense green emission is confirmed under UV irradiation (see Figs. 1(b) and 2(a)). First, we evaluated the fundamental optical properties of CsPbBr$_3$/Cs$_4$PbBr$_6$. Figure 2(a) shows the optical absorption and the PL spectra. The absorption spectrum exhibits a steep onset at 2.35 eV and a peak at 2.42 eV. The PL peak energy agrees well with the absorption onset. This suggests that the green PL originates from the band-to-band optical transition of the embedded CsPbBr$_3$, while the bandgap energy Cs$_4$PbBr$_6$ is ~4 eV [35]. The PL spectrum of a CsPbBr$_3$ bulk crystal is also plotted in the same figure for reference. Note that the slight

PL blueshift of the composite crystal with respect to the CsPbBr$_3$ bulk crystal is attributable to the quantum confinement of the embedded CsPbBr$_3$ nanostructures and the small photon reabsorption effect [25].

Figure 2(b) shows the typical PL external quantum efficiency (EQE) as a function of excitation power density using an integrating sphere [12]. In this measurement, the maximum EQE was 84%, which is below unity probably due to defects and impurities. Note that the EQE shows large sample-to-sample dependence and ranges from 50~100%. We record the maximum value of EQE > 90%. Considering the much higher EQE of CsPbBr$_3$/Cs$_4$PbBr$_6$ than CsPbBr$_3$ bulk crystals (~2% at the maximum; see Supplemental Material [39]), we can conclude that nonradiative losses are strongly suppressed by embedding CsPbBr$_3$ nanostructures into a Cs$_4$PbBr$_6$ host crystal. Furthermore, the EQE decreases with an increase in the excitation power density. This result suggests that nonlinear recombination processes, such as Auger recombination and exciton–exciton annihilation (EEA), become dominant due to charge accumulation under intense photoexcitation [51]. The recombination processes are discussed later based on time-resolved PL measurements. We confirmed that the PL quench at high excitation power densities does not originate from persistent damage.

Figure 2(c) displays the typical behaviors of the PL intensity under continuous photoexcitation. The PL intensity of the CsPbBr$_3$/Cs$_4$PbBr$_6$ is more stable compared with the CsPbBr$_3$ bulk counterpart. Light-induced change in the PL intensity is well known in halide perovskites and has been discussed in terms of photo-induced ion migration and carrier trapping [52]. The high photostability of the CsPbBr$_3$/Cs$_4$PbBr$_6$ is attributable to the unique composite structure, i.e., nanostructures protected by the host material.

It is essential to reveal the recombination dynamics for thorough understanding of the PL process. Figure 3(a) shows the streak image under excitation at 10 μJ/cm$^2$. The horizontal and vertical axes are the emission energy and the delay time, respectively. Several time-resolved PL spectra are shown in Figure 3(b). The PL peak locates at 2.39 eV just after excitation and exhibits a slight redshift (≈16 meV) within a few tens of nanoseconds. As shown in the inset of Figure 3(c), the PL decay possesses fast (a few nanoseconds) and slow (~150 ns) decay components. The decay time of the fast component is similar to the exciton lifetime reported in CsPbBr$_3$ quantum dots [53]. Considering the possible size distribution of the CsPbBr$_3$ nanostructure, the slow component is attributable to the large nanostructures. This explanation is consistent with the PL redshift that occurs up to about 50 ns.

We also examined the excitation-fluence dependence of the PL dynamics monitored at 2.39 eV, as shown in Figure 3(c). The abovementioned fast component appears as the excitation fluence is increased. The PL intensity just after excitation, $I_{PL}^{t=0}$, and the effective lifetime, $\tau_{\text{eff}}$, are summarized in Figure 3(d) as a function of the incident photon flux $N_{ph}$. Here, $\tau_{\text{eff}}$ is defined by the weighted average of the three lifetimes estimated from the fit to a triple-exponential function. The $I_{PL}^{t=0}$ values follow a power-law dependence on the excitation fluence with an exponent of 1.1. If exciton recombination dominates the radiative recombination process, $I_{PL}^{t=0}$ should exhibit linear dependence on the excitation fluence, as observed in quantum dots. On the other hand, a square dependence is usually observed in bulk semiconductors where bimolecular recombination is dominant [23]. Our result for $I_{PL}^{t=0}$ is close to the excitonic case, suggesting that our Cs$_4$PbBr$_6$ crystal contains quantum-sized CsPbBr$_3$ nanostructures.

The $\tau_{\text{eff}}$ values are reduced at high excitation densities. This reduction in the lifetime is attributable to the nonradiative recombination due to EEA, and is consistent with the low EQE observed under strong excitation (Figure 2(b)). We used the following rate equation to explain this behavior:

$$\frac{dN}{dt} = -AN - BN^2 \qquad (1)$$

where $N$, $A$, and $B$ are the exciton density, the exciton recombination rate, and the EEA rate coefficient, respectively. According to Eq. (1), the effective lifetime is formally written as $\tau_{\text{eff}} = \frac{1}{(A+BN)}$, which agrees well with the experimental results (see broken curve in Figure 3(d)). This result is consistent with the abovementioned excitonic recombination.

Anti-Stokes PL is essential for the realization of optical refrigeration. Therefore, we measured PL excitation (PLE) spectra. Figure 4(a) shows the PLE map. The vertical and horizontal axes are the excitation and the emission energies, respectively. The white diagonal line corresponds to the scattering signal of the excitation light. On the right side of the diagonal line, anti-Stokes PL can be clearly observed. To display the anti-Stokes PL more clearly, we plot several PL spectra under different excitation photon energies $E_{ex}$ in Figure 4(b). Under low-energy excitation below the emission peak energy $E_{PL}$, the PL peak appears at the high-energy side of the excitation light. Moreover, the spectral shape is independent of the excitation energy, which means that the PL linewidth is governed by homogeneous broadening. To eliminate the possibility that the anti-Stokes PL originates from multiphoton absorption, we measured the excitation-density dependence of the PL intensity under 2.34-eV excitation (see the inset of Figure 4(c)). The observed linear dependence indicates that one-photon absorption is dominant.

Next, we focus on the optical absorption at energies below $E_{PL}$. Intrinsic optical absorption below exciton energy occurs owing to electron-phonon interactions. One example is the LO-phonon replica, as observed in CdS nanobelt [8]. However, no phonon replica signals are observed in PL and optical absorption spectra of CsPbBr$_3$/Cs$_4$PbBr$_6$. Instead, Urbach tail absorption is dominant and the spectral shape is written by [54]

$$\alpha(E) = \alpha_0 \exp\left(\frac{E - E_0}{E_U}\right) \qquad (2)$$

where $E_U$ and $E_0$ are the Urbach energy and the Urbach focus energy, respectively. The Urbach energy is contributed both from intrinsic electron–phonon interactions and extrinsic effects such as lattice distortion by defects and impurities. The former contribution realizes the photoexcitation below the absorption-edge energy via phonon-assisted process, resulting in the anti-Stokes PL. The latter part causes an inhomogeneous broadening of the PL spectrum and constitutes a limiting factor for optical refrigeration. We evaluated the Urbach energy of CsPbBr$_3$/Cs$_4$PbBr$_6$ from the PLE spectrum shown in Figure 4(c) using Eq. (2). We obtained $E_U = 16.6$ meV, which agrees well with that estimated from the PL spectrum using the van Roosbroeck–Shockley relation (16.1 meV; Supplemental Material [39]). This value is close to those reported in CsPbBr$_3$ bulk crystals (19 meV [55]) and nanocrystals (~13 meV [56]). Note that the Urbach energy of CsPbBr$_3$ is considerably larger than those of conventional inorganic semiconductors such as GaAs (~5 meV [57]) and CdTe (~8 meV [58]).

It is meaningful to estimate the minimum EQE required for cooling, $\eta_{ext}$, which can be expressed in terms of $E_{ex}$, $E_{PL}$, and the fraction of excitation photons that are engaged in cooling, $\eta_{abs}$ [5]:

$$\eta_{ext} = \frac{1}{\eta_{abs}} \frac{E_{ex}}{E_{PL}}. \tag{3}$$

If we assume that we can excite at $E_{ex} = E_{PL} - kE_U$, we obtain $\eta_{ext} \cong 1 - \frac{kE_U}{E_{PL}}$. Here $k$ is a constant (presumably in the range of about 2 to 5; see Supplemental Material [39]) that depends on the parasitic photoabsorption in the sample. Therefore, a larger intrinsic Urbach energy allows photoexciting at lower energies, enabling a larger anti-Stokes shift, $E_{PL} - E_{ex}$, which enhances the cooling gain per absorbed photon. We also assumed the ideal case $\eta_{abs} = 1$. We estimated the minimum EQE required for cooling for the case of $k = 4$ and obtained $\eta_{ext} = 97.2\%$ for CsPbBr$_3$/Cs$_4$PbBr$_6$ according to Eq. (3) (c.f. GaAs: 98.6%, InP: 98.1%, and GaN: 98.5% calculated based on the $E_U$ in the literature [59]). Because the highest reported EQE exceeds 97% [35], the above estimation suggests a high possibility of realizing optical cooling using CsPbBr$_3$/Cs$_4$PbBr$_6$.

In general, the Urbach tail is determined both by intrinsic phonon-assisted process and extrinsic structural disorder. To elucidate the contribution of the intrinsic phonon-assisted process, we conducted temperature-dependent PL measurements. Figure 5(a) shows PL spectra from 70 to 250 K at high-energy (2.53 eV) and low-energy (2.32 eV) excitations. The PL spectrum shape is almost independent of the excitation photon energy even at low temperatures, and this means that homogeneous broadening is dominant in the whole measured temperature range. We also find that the PL peak shows a redshift and the PL linewidth becomes narrower as the temperature decreases. Here, the PL linewidth was characterized using the full width at half-maximum (FWHM). Furthermore, even at low temperatures near 70 K, the PL spectrum is composed of a single peak [39].

Figure 5(b) shows the temperature dependence of the PLE spectrum. The absorption edge becomes steeper at lower temperatures. The Urbach energy at each temperature was obtained using Eq. (2). We summarize the temperature dependences of $E_U$, $E_{PL}$, and the PL FWHM in Figure 5(c). The Urbach energy derived from the PL spectrum using the van Roosbroeck–Shockley relation is also plotted in this figure [39]. We conducted curve fitting for $E_U$, $E_{PL}$, and the PL FWHM as shown by the solid and dashed curves (see Supplemental Material for details regarding $E_{PL}$ and the PL FWHM [39]).

The temperature dependence of the Urbach energy can be analyzed by the following equation [54,60]:

$$E_U(T) = \frac{\hbar\omega_{\text{ph}}}{2\sigma_0}\left(\coth\left(\frac{\hbar\omega_{\text{ph}}}{2k_B T}\right) + X\right) \qquad (4)$$

where $\sigma_0$ is the steepness parameter, independent of the temperature $T$, and $\hbar\omega_{\text{ph}}$ is the averaged energy of the phonons that contribute to the Urbach tail. $X$ represents the contribution of structural disorder. According to Eq. (4), $E_U(T)$ approaches $\frac{k_B T}{\sigma_0}$ as $T \to \infty$. To estimate $\sigma_0$, we used the high-temperature data (above 200 K) and obtained $\sigma_0 = 1.75 \pm 0.06$, which is the same as that of CsPbBr$_3$ bulk crystals [38]. It should be emphasized that this value is smaller than those of conventional inorganic semiconductors (cf. 2.2 for CdS [58] and 5.0 for GaAs [59]). The solid curve in the upper panel of Figure 5(c) is the fit to Eq. (4), where we used $\sigma_0 = 1.75$. We obtained $\hbar\omega_{\text{ph}} = 3\ (\pm 2)$ meV and $X < 0.6$. The estimated $\hbar\omega_{\text{ph}}$ is close to the lowest LO phonon energies (~4 meV) reported in CsPbBr$_3$ [61–64]. Note that the structural disorder, $X$, is negligibly small compared with the intrinsic phonon contribution at least above 100 K. This result evidences that the phonon-assisted process dominates the Urbach tail of absorption and anti-Stokes PL at around room temperature.

Finally, we discuss the potentially useful functions of halide perovskite nanostructures for optical cooling from the viewpoint of electron–phonon interactions. Let us consider a steepness parameter, which is a metric of electron-phonon interactions. According to Eq. (4), a small $\sigma_0$ is essential to realize a large Urbach energy. On the other hand, according to the seminal work on self-trapped excitons (STE) by Toyozawa [58], STEs are formed when $\sigma_0$ is below a critical value (1.64 from theoretical prediction), which has been confirmed in various materials, including alkali halides, organic semiconductors, and inorganic semiconductors [58]. STE states give rise to a large PL Stokes shift due to lattice relaxation, resulting in the reduction of anti-Stokes PL component, which is detrimental to optical cooling. Our estimated steepness parameter ($\sigma_0 = 1.75$) suggests that CsPbBr$_3$ nanostructure is a promising material for optical cooling from the viewpoint of electron-phonon interactions because it provides a large Urbach energy without falling into the STE state.

On the other hand, there are several challenges that must be overcome for practical perovskite-based optical refrigerators. The degradation of halide perovskites by light and air exposure is still a problem to be solved, especially in their nanostructures. Furthermore, it is also necessary to increase the absorbing volume to enhance the cooling power to make a practical device, while the optical cooling has previously been reported only for nano- or micro-sized perovskites [13,16]. It is generally difficult to achieve both a large volume and high PL efficiency because the nonradiative recombination rate of the nanoparticles is usually smaller than that of the large-sized bulk counterpart. These problems may be solved by protecting the perovskite nanostructures with host materials, such as CsPbBr$_3$/Cs$_4$PbBr$_6$. As for CsPbBr$_3$/Cs4PbBr$_6$, the highest reported EQE is over 97% [35], which is at a level where the optical cooling is expected. If we can control the density and the size of CsPbBr$_3$ nanostructures dispersed in the host material, CsPbBr$_3$/Cs$_4$PbBr$_6$ can be a promising candidate for semiconductor optical cooling. It should also be mentioned that the anti-Stokes optical cooling using Urbach tail absorption is a self-limited process, because the phonon density is reduced as decreasing temperature. Therefore, it is necessary to understand the detailed mechanisms of the anti-Stokes PL at low temperatures to evaluate the lowest achievable temperature.

In conclusion, we observed efficient anti-Stokes PL from CsPbBr$_3$ nanostructures embedded in a host Cs$_4$PbBr$_6$ crystal. We reported the high photostability of CsPbBr$_3$/Cs$_4$PbBr$_6$, which is probably due to its unique guest/host structure. We have presented the temperature dependence of the Urbach energy determined from the PLE spectra and revealed that intrinsic phonon-assisted process dominates the Urbach tail of absorption and anti-Stokes PL processes. Our spectroscopic studies suggest the potential of CsPbBr$_3$ nanostructures for optical cooling from the viewpoint of electron-phonon interactions.

**Acknowledgments**

The authors would like to thank Prof. S. Chichibu and Prof. K. Oto for their cooperation during this work. Part of this work was supported by JST-CREST (Grant No. JPMJCR16N3), JSPS KAKENHI (Grant No. JP19K03683, JP19H05465), the Canon Foundation, and the Chiba Iodine Resource Innovation Center.

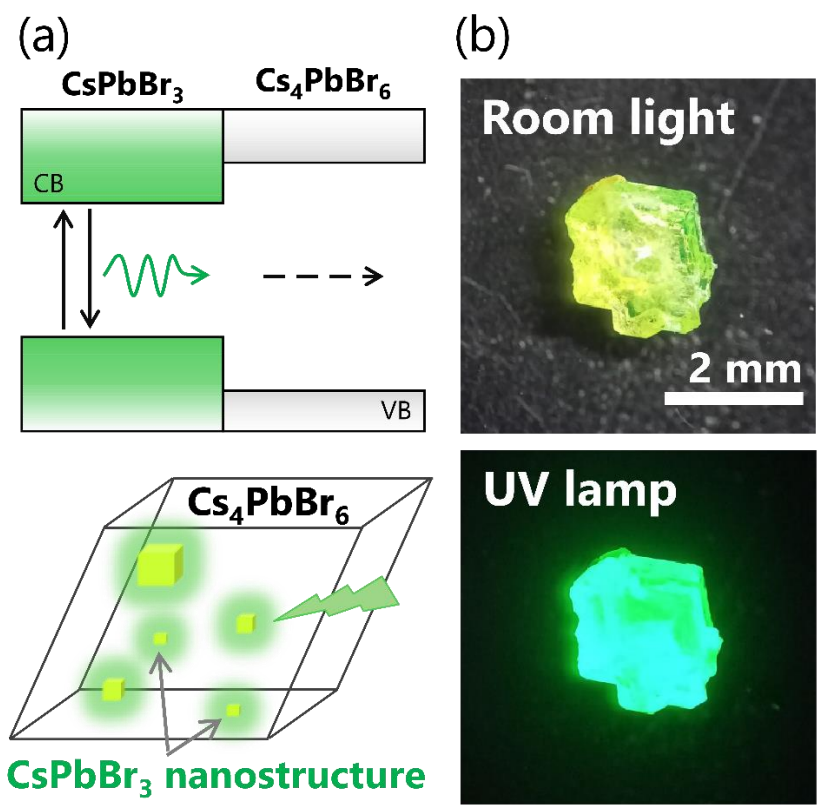

**Figure 1.** (a) Energy diagram of $CsPbBr_3$ and $Cs_4PbBr_6$ (top panel) and schematic of $CsPbBr_3$ nanostructures in a $Cs_4PbBr_6$ crystal (bottom panel). (b) Photographs of a $CsPbBr_3$/$Cs_4PbBr_6$ crystal under room light and UV lamp excitation.

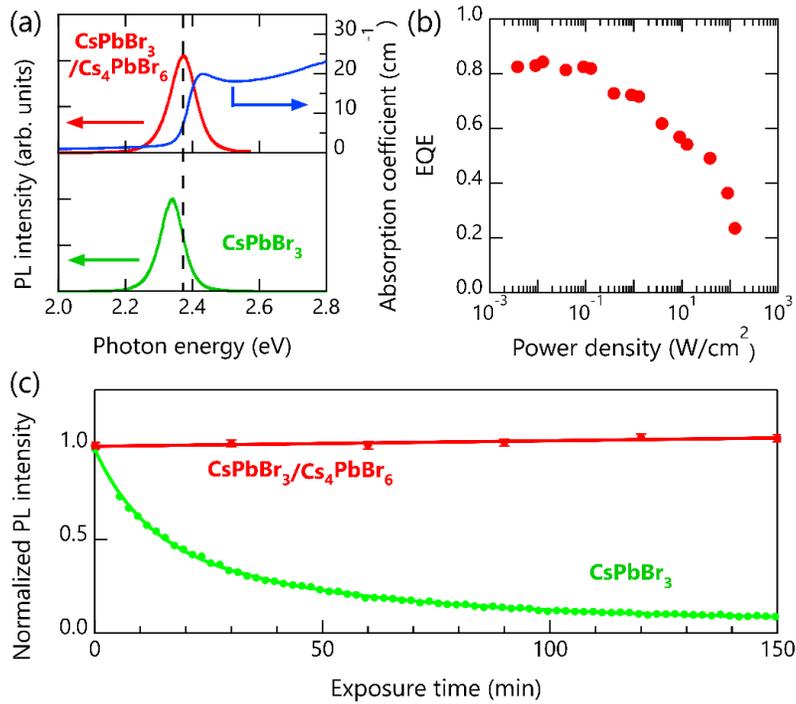

**Figure 2.** (a) Optical absorption and PL spectra of $CsPbBr_3/Cs_4PbBr_6$ (top panel) and $CsPbBr_3$ bulk crystals (bottom panel). (b) EQE as a function of the excitation power density for the $CsPbBr_3/Cs_4PbBr_6$ crystal. Excitation photon energy was 3.1 eV. (c) Dependence of the PL intensity on the exposure time for the $CsPbBr_3/Cs_4PbBr_6$ (red data) and the $CsPbBr_3$ bulk crystal (green data) under continuous-wave photoexcitation in air at room temperature (excitation intensity: 2 kW/cm$^2$; photon energy: 3.0 eV). The PL intensities are normalized to the values at the beginning of the experiment. Solid curves are eye guides.

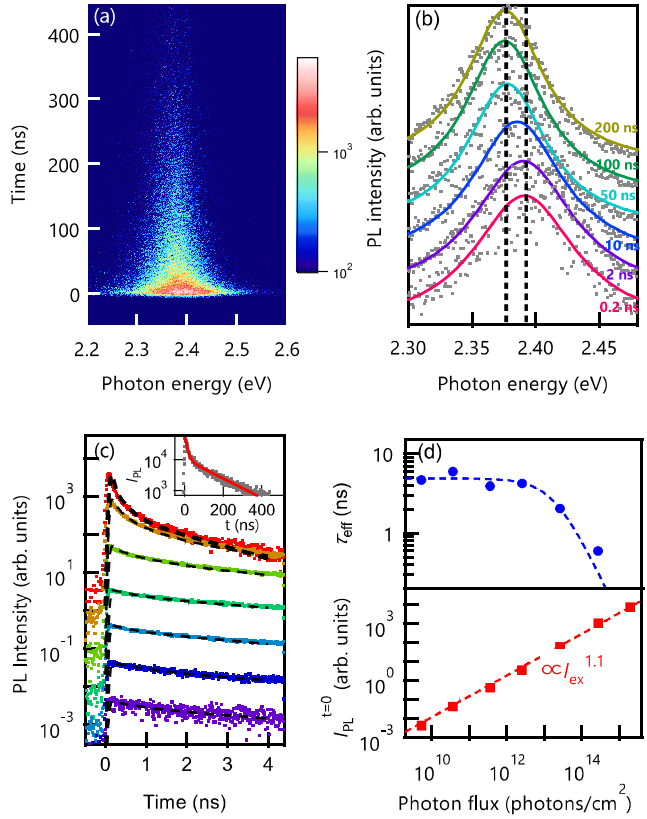

**Figure 3.** (a) 2D streak data of $CsPbBr_3/Cs_4PbBr_6$ under excitation at 2.99 eV (10 μJ/cm$^2$). (b) Time-resolved PL spectra at different times after excitation. Solid curves are fits to a pseudo-Voigt function. (c) Excitation fluence dependence of the PL decay profile monitored at 2.39 eV (2 nJ/cm$^2$ ~ 1 μJ/cm$^2$). The broken curves are fits to a triple-exponential function. The inset shows the PL dynamics on a longer time scale. (d) Effective lifetime and PL intensity just after excitation as a function of $N_{ph}$. We fit $\tau_{eff}$ data to function $1/(a + bN_{ph})$.

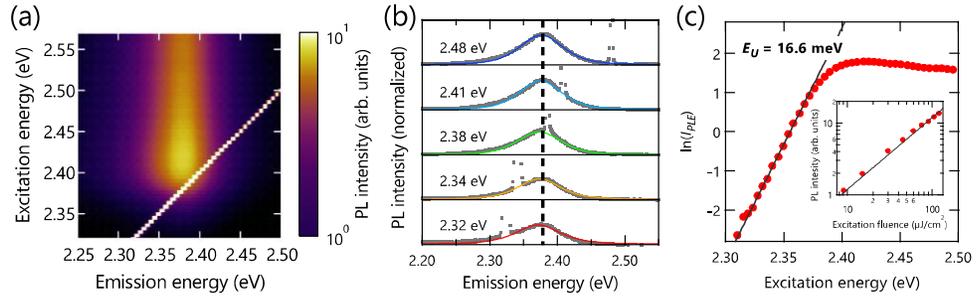

**Figure 4.** (a) PLE map of CsPbBr$_3$/Cs$_4$PbBr$_6$. (b) PL spectra under different excitation photon energies. Solid curves are fits to a pseudo-Voigt function. (c) Integrated PL intensity ($I_{PLE}$) as a function of excitation photon energy ($E_{ex}$). The solid line is a fit to Eq. (2). The inset shows the excitation-fluence dependence of the integrated anti-Stokes PL intensity under 2.34-eV excitation. The solid line indicates the linear dependence.

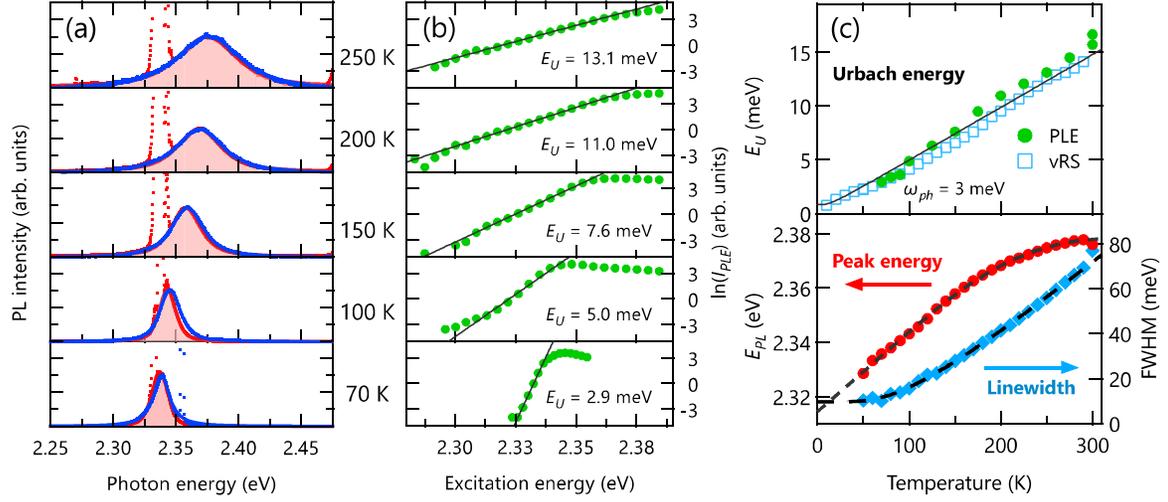

**Figure 5.** (a) PL and (b) PLE spectra at different temperatures. Blue and red curves in (a) represent the spectra obtained under excitation at 2.53 eV and 2.33 eV, respectively. The solid lines in (b) are fits to Eq (2), and the evaluated Urbach energy $E_U$ is provided in each figure. (c) The closed circles show $E_U$, $E_{PL}$, and the PL linewidth as a function of the sample temperature. The Urbach energies derived from the PL spectrum using the van Roosbroeck–Shockley relation are shown by the open squares. The solid curve in the upper panel is a fit to Eq. (4), from which we obtained $\hbar\omega_{\text{ph}} = 3$ meV. The dashed curves in the middle and lower panels are fits to Eq. (S3) and (S4) in Supplemental Material [38].

20